\documentclass{article}

\usepackage[preprint]{neurips_2026}

\usepackage[utf8]{inputenc} % allow utf-8 input
\usepackage[T1]{fontenc}    % use 8-bit T1 fonts
\usepackage{hyperref}       % hyperlinks
\usepackage{url}            % simple URL typesetting
\usepackage{booktabs}       % professional-quality tables
\usepackage{amsfonts}       % blackboard math symbols
\usepackage{nicefrac}       % compact symbols for 1/2, etc.
\usepackage{microtype}      % microtypography
\usepackage{xcolor}         % colors
\usepackage{graphicx}
\usepackage{multirow}

\title{Stage-adaptive audio diffusion modeling}

\author{%
  Xuanhao Zhang \\
  China Pharmaceutical University\\
  \texttt{2020230870@stu.cpu.edu.cn} 
  \and
  \textbf{Chang Li} \\
  University of Science and Technology of China \\
  \texttt{lc\_lca@mail.ustc.edu.cn} \\
}

\begin{document}
\maketitle
\begin{abstract}
Recent progress in diffusion-based audio generation and restoration has substantially improved performance across heterogeneous conditioning regimes, including text-conditioned audio generation and audio-conditioned super-resolution. However, training audio diffusion models remains computationally expensive, and most existing pipelines still rely on static optimization recipes that treat the relative importance of training signals as fixed throughout learning. In this work, we argue that a major source of inefficiency lies in the evolving balance between semantic acquisition and generation-oriented refinement. Early training places stronger emphasis on acquiring condition-aligned semantic structure and coarse global organization, whereas later training increasingly emphasizes temporal consistency, perceptual fidelity, and fine-detail refinement. To characterize this evolving balance, we introduce a progress-based regime variable derived from the training-time slope of an SSL-space discrepancy, which measures semantic progress during training. Based on this signal, we develop three complementary stage-aware mechanisms: decayed SSL guidance for early semantic bootstrapping, self-adaptive timestep sampling driven by the regime variable, and structure-aware regularization activated from convergent grouped organization in parameter space. We evaluate these mechanisms on text-conditioned audio generation and audio-conditioned super-resolution. Across both settings, the proposed stage-aware strategies improve convergence behavior and yield gains on the primary generation and spectral reconstruction metrics over standard static baselines. These results support the view that efficient audio diffusion training can benefit from treating external guidance, internal organization, and optimization emphasis as stage-dependent components rather than fixed ingredients.
\end{abstract}

\section{Introduction}

Diffusion-based audio models have achieved strong performance across heterogeneous conditioning regimes, including text-conditioned audio generation~\cite{liu2023audioldm} and audio-conditioned restoration such as super-resolution~\cite{liu2024audiosr,li2025audio}. However, high-fidelity audio diffusion training remains computationally expensive, and practical improvements are still often driven by scaling data, model size, or optimization budget rather than by improving the training process itself. This motivates a shift from resource-driven scaling toward training-efficient audio diffusion modeling.

We argue that a major limitation of current audio diffusion pipelines is their reliance on a largely static training recipe. During training, a conditional audio diffusion model must gradually acquire condition-aligned semantic structure, develop long-range temporal organization, and synthesize fine-grained perceptual details. These abilities do not emerge at the same pace, suggesting that the relative usefulness of different training signals may change over optimization. Early training places stronger emphasis on semantic grounding and coarse global organization from the conditioning signal, while later training increasingly emphasizes temporal consistency, perceptual fidelity, and local detail refinement. Treating the entire process as a stationary optimization problem may therefore lead to inefficient allocation of training signals.

This view is consistent with recent evidence from efficient diffusion training: representation alignment can substantially accelerate diffusion transformer training~\cite{yu2024representation}, block-wise representations in diffusion transformers evolve non-uniformly during training~\cite{yang2026diversedit}, and adaptive timestep sampling improves convergence by allocating training effort to more informative timesteps~\cite{kim2025adaptive}. Together, these findings motivate the view that semantic guidance, internal structure, and timestep allocation may need to evolve during training rather than remain fixed. Yet existing methods typically study these factors separately, without a unified mechanism for adapting them to the model's training stage.

To address this gap, we introduce a stage-aware perspective on efficient audio diffusion training. The central idea is to track the evolving balance between semantic acquisition and generation-oriented refinement, and to use this signal to adapt the training process. Specifically, we define a regime variable from the training-time slope of an SSL-space discrepancy between the predicted and target clean signals. Larger progress values indicate rapid improvement in semantic representation space, whereas smaller values indicate that semantic progress is saturating and that training may benefit from greater emphasis on refinement-oriented signals.

Based on this regime variable, we develop three complementary stage-aware mechanisms. First, we introduce decayed SSL guidance from a frozen audio self-supervised encoder, providing stronger semantic support early in training and gradually reducing this external guidance as the model internalizes semantic structure. Second, we propose regime-driven adaptive timestep sampling, which reallocates training effort as the dominant optimization bottleneck shifts from semantic acquisition toward detail refinement. Third, we impose structure-aware regularization over transformer blocks based on emergent grouped organization in parameter space, activating this coupling only when internal structure becomes stable and informative. These mechanisms instantiate the same principle: training signals should evolve with the model's stage rather than remain fixed.

We evaluate the proposed perspective on two representative settings: text-conditioned audio generation and audio-conditioned super-resolution. Across both tasks, stage-aware training improves convergence behavior and yields gains on the primary generation and spectral reconstruction metrics over static baselines. Further analyses of semantic progress, timestep preference, and block-wise structural evolution suggest that efficient audio diffusion training can be better understood as a progressive process in which external guidance, internal organization, and optimization emphasis co-evolve.

Our contributions are summarized as follows:
\begin{itemize}
    \item We propose a stage-aware perspective on efficient audio diffusion training, centered on the evolving balance between semantic acquisition and generation-oriented refinement.

    \item We introduce a regime variable derived from the training-time slope of an SSL-space discrepancy to characterize semantic progress and guide training-stage adaptation as the dominant optimization bottleneck changes over time.

    \item We develop three stage-aware mechanisms: decayed SSL guidance, regime-driven adaptive timestep sampling, and structure-aware regularization over transformer blocks, which respectively adapt semantic support, timestep emphasis, and structural coupling.

    \item We validate the approach on text-conditioned audio generation and audio-conditioned super-resolution, with analyses of convergence, semantic dynamics, timestep allocation, and structural evolution, demonstrating improvements on text-to-audio quality and spectral reconstruction metrics over static training baselines.
\end{itemize}
% =========================
% Method
% =========================
\section{Method}
\label{sec:method}

\subsection{Overview and Preliminaries}
\label{sec:overview_preliminaries}

We study efficient audio diffusion training under heterogeneous conditioning regimes. Our starting point is that conditional audio diffusion optimization is not well served by a fully static training recipe: although the diffusion objective remains fixed, the relative usefulness of different training signals can change over the course of learning. Early training places stronger emphasis on condition-aligned semantic acquisition and coarse global organization, whereas later training increasingly emphasizes temporal consistency, perceptual fidelity, and fine-detail refinement. This motivates a stage-aware view of audio diffusion training, where semantic support, structural coupling, and optimization emphasis are adapted according to the model's evolving training state.

To make this tradeoff explicit, we introduce a semantic progress signal defined in a frozen SSL representation space. Let $\theta_k$ denote the model parameters at training iteration $k$, and let $\hat{x}_0^{(k)}$ denote the model prediction of the clean target under the adopted diffusion parameterization. We quantify semantic progress by comparing the predicted and ground-truth clean signals in SSL space through a frozen encoder $E_{\mathrm{ssl}}(\cdot)$ and a discrepancy measure $D(\cdot,\cdot)$. Every $\Delta k$ steps, we sample a mini-batch $\mathcal{B}_k$ and evaluate a stabilized SSL-space discrepancy:
\begin{equation}
\widetilde{\mathcal{L}}_{\mathrm{ssl}}^{(k)}
=
\frac{1}{|\mathcal{B}_k|\,|\mathcal{R}|}
\sum_{x_0 \in \mathcal{B}_k}
\sum_{r \in \mathcal{R}}
D\!\left(
E_{\mathrm{ssl}}\!\left(r(\hat{x}_0^{(k)})\right),
E_{\mathrm{ssl}}\!\left(r(x_0)\right)
\right),
\end{equation}
where $\mathcal{R}$ denotes a small set of deterministic bilinear smoothing views applied symmetrically before SSL feature extraction, yielding a stabilized discrepancy estimate for the subsequent training-time slope computation. In all experiments, we set $\Delta k = 500$.

We estimate semantic progress from the local trend of recent discrepancy observations. Let $\mathcal{W}_k=\{(k_i,\widetilde{\mathcal{L}}_{\mathrm{ssl}}^{(k_i)})\}_{i=1}^{m}$ denote the most recent $m$ observations up to step $k$. We fit a local linear trend $\widetilde{\mathcal{L}}_{\mathrm{ssl}}^{(k_i)}\approx a_k k_i+b_k$ over $\mathcal{W}_k$ and define the regime variable as $g_k=-a_k$. Equivalently, $g_k$ approximates the negative training-time slope of the stabilized SSL discrepancy. Thus, a larger $g_k$ indicates faster progress in SSL representation space, whereas a smaller $g_k$ indicates that semantic improvement is saturating. In our framework, $g_k$ is used purely as a monitoring signal for stage-aware control and is not backpropagated through the diffusion objective.

This variable serves as a compact indicator of the evolving balance between semantic acquisition and generation-oriented refinement. Larger values of $g_k$ correspond to rapid improvement in SSL representation space, suggesting that semantic grounding and condition alignment remain highly responsive to external support. As $g_k$ decreases, semantic progress becomes less pronounced, and training can place relatively greater emphasis on temporal consistency, perceptual fidelity, and fine-detail refinement. In our framework, $g_k$ therefore acts as a shared progress signal that modulates the strength of external semantic guidance, the activation of internal structural coupling, and the allocation of optimization effort across diffusion timesteps.

We consider two representative audio diffusion settings: text-conditioned audio generation and audio-conditioned super-resolution. In both cases, a Diffusion Transformer (DiT) denoiser $f_\theta$ is trained in latent space. Let $x_0$ denote a clean target audio sample and let $z_0$ be its latent representation. Given a sampled timestep $t \sim p(t)$, the forward process produces a corrupted latent $z_t$, and the denoiser predicts the corresponding training target under conditioning input $c$, which may be text or low-frequency audio depending on the task. The base training objective is
\begin{equation}
\mathcal{L}_{\mathrm{diff}}(\theta)
=
\mathbb{E}_{z_0,\,t,\,\epsilon}
\Big[
\ell\!\left(f_\theta(z_t,t,c),\,u\right)
\Big],
\end{equation}
where $u$ denotes the diffusion target under the adopted parameterization and $\ell(\cdot,\cdot)$ is the training loss, instantiated as squared error in our experiments.

Building on this objective, we develop three stage-aware components organized around $g_k$. First, we introduce decayed SSL guidance from a frozen audio self-supervised encoder, providing stronger semantic support when SSL-space progress is rapid and gradually reducing this external guidance as semantic acquisition stabilizes. Second, we employ self-adaptive timestep sampling to reallocate optimization effort according to the evolving progress signal, increasing the relative emphasis on refinement-oriented timesteps as semantic progress becomes less pronounced. Third, we impose structure-aware regularization over transformer blocks based on emergent grouped organization in parameter space, activating this coupling once internal structure becomes sufficiently stable and informative. Taken together, these components use the same progress signal to adapt semantic support, timestep emphasis, and structural coupling throughout training, aligning the training recipe with the model's evolving optimization state.
\subsection{Decayed SSL Guidance}
\label{sec:ssl_guidance}

We first introduce an external semantic prior to support early semantic acquisition. In our framework, larger values of $g_k$ indicate rapid improvement in SSL representation space, suggesting that the denoiser can benefit from additional semantic structure while learning to align noisy diffusion targets with the conditioning signal. We therefore inject features from a frozen pretrained audio encoder into the DiT through cross-attention, providing an informative semantic scaffold during optimization.

The SSL condition is designed as a transient source of guidance rather than a permanent dependency. As $g_k$ decreases, SSL-space progress becomes less pronounced, and training can place relatively greater emphasis on temporal consistency and perceptual refinement. We therefore progressively decay the SSL condition, encouraging the denoiser to rely less on external semantic features while retaining their early optimization benefit. Decayed SSL guidance thus adapts external semantic support according to the evolving progress signal.

Given a clean audio sample $x$, we extract SSL features using a frozen pretrained encoder $E_{\mathrm{ssl}}$ and map them into the conditioning space of the denoiser through a lightweight projector. The resulting SSL representation is injected into transformer blocks as an auxiliary cross-attention context. To avoid imposing a fixed external semantic prior throughout optimization, we make this SSL condition stage-dependent and denote the effective injected context at training step $k$ by $\widetilde{C}_{\mathrm{ssl},k}$.

To progressively transfer semantic responsibility from the external condition to the model itself, we apply a decay schedule
\begin{equation}
\gamma_k
=
\max\left(0,\ 1-\frac{k}{\rho_{\mathrm{ssl}} K_{\mathrm{tot}}}\right),
\qquad \rho_{\mathrm{ssl}} \in (0,1],
\end{equation}
where $K_{\mathrm{tot}}$ is the total number of training steps and $\rho_{\mathrm{ssl}} K_{\mathrm{tot}}$ specifies when the SSL guidance fully vanishes. At each training step, we mask the projected SSL representation in both the time and frequency dimensions according to the current decay ratio. Concretely, let $C_{\mathrm{ssl}}$ denote the projected SSL representation and let $M_k$ denote the corresponding binary time-frequency mask generated with active ratio controlled by $\gamma_k$. The effective injected SSL condition is then given by
\begin{equation}
\widetilde{C}_{\mathrm{ssl},k} = C_{\mathrm{ssl}} \odot M_k.
\end{equation}
% Figure~\ref{fig:masked_spectrograms} illustrates this progressive masking process. As the masking ratio increases, larger portions of the SSL-derived time-frequency structure are removed, providing a qualitative view of how externally injected semantic information is gradually reduced during training.

The denoiser is then trained with
\begin{equation}
L_{\mathrm{diff}}^{(k)}
=
\mathbb{E}_{z_0,\, t,\, \epsilon}
\left[
\|u - f_\theta(z_t, t, c, \widetilde{C}_{\mathrm{ssl},k})\|_2^2
\right].
\end{equation}
Early in training, this design provides a high-coverage semantic scaffold that allows the model to quickly exploit informative SSL features. Later, the progressive masking forces the model to internalize and reconstruct the same semantic capability without continued reliance on the external prior. In this way, decayed SSL guidance accelerates semantic acquisition while explicitly encouraging the transition from externally provided semantics to model-owned semantic understanding.
\subsection{Self-Adaptive Timestep Sampling}
\label{sec:adaptive_timestep}

The previous two mechanisms improve training efficiency by strengthening semantic acquisition in the early stage and exploiting emergent internal organization once it becomes reliable. However, standard diffusion training still relies on a fixed timestep sampling policy throughout optimization, implicitly assuming that supervision should be allocated over noise levels in the same way at all stages of learning. This is suboptimal. In practice, the optimization bottleneck of audio diffusion models is stage-dependent: early training is primarily concerned with forming coarse semantic and structural organization, whereas later training increasingly shifts toward perceptual refinement and fine-detail reconstruction. As a result, the usefulness of different timesteps is not static over training, and timestep allocation should adapt accordingly. This view is also consistent with recent work showing that non-uniform timestep sampling can substantially improve diffusion training efficiency~\cite{zheng2024non}.

Conceptually, timestep sampling can be viewed as an allocation of training budget over the diffusion trajectory. Ideally, the sampling distribution at iteration $k$ should place more mass on timesteps with higher marginal optimization utility,
\begin{equation}
p_k^\star(t) \propto \mathcal{U}_k(t),
\end{equation}
where $\mathcal{U}_k(t)$ denotes the expected optimization benefit of sampling timestep $t$ at the current stage of training. While this formulation captures the desired principle, directly estimating $\mathcal{U}_k(t)$ online is impractical in large-scale diffusion training. We therefore seek a low-overhead surrogate that captures stage-wise optimization needs through the semantic progress variable introduced in Section~\ref{sec:overview_preliminaries}.

Let $g_k$ denote the semantic progress signal. Since this quantity can be negative in practice, we first map it to the positive real line through an exponential transform, and then convert it into a stage-dependent mode parameter $\mu_k \in (0,1]$. This mapping is designed to satisfy three boundary conditions: when semantic progress becomes very large, the resulting timestep distribution should peak near $\tau=0$; when semantic progress approaches zero, the mode should shift toward $\tau=1$; and when the transformed progress signal is close to one, the distribution should peak near $\tau=0.5$. Here $\tau=t/T \in [0,1]$ denotes the normalized timestep, with $T$ the total number of diffusion steps.

We instantiate the sampling distribution as a Beta family whose mode is controlled by $\mu_k$:
\begin{equation}
p_k(\tau)=\mathrm{Beta}\!\left(\tau;\alpha_k,\beta_k\right),
\end{equation}
where the shape parameters $\alpha_k$ and $\beta_k$ are chosen so that the mode of $p_k(\tau)$ is exactly $\mu_k$, while a concentration parameter controls how sharply probability mass is concentrated around that mode. In this way, the timestep sampler evolves smoothly with the semantic progress regime: when semantic progress remains strong, the sampler favors timestep regions associated with coarse organization; when semantic progress saturates, probability mass is gradually reallocated toward regions that better support generation-oriented refinement.

This design can be interpreted as a tractable approximation to the ideal utility-based sampler in Eq.~(15). Instead of explicitly estimating the optimization utility of each timestep, which would be costly and unstable, we use the observed semantic progress dynamics of the model as a global proxy for stage-wise utility variation. In this way, timestep sampling becomes a dynamic function of the evolving optimization regime rather than a fixed prior imposed throughout training.
\subsection{Structure-Aware Regularization}
\label{sec:structure_regularization}

Decayed SSL guidance adapts the amount of external semantic support, but it does not directly account for another source of training dynamics: the internal organization of DiT blocks also evolves over optimization. Transformer blocks need not contribute independently or uniformly across depth; as training proceeds, their parameter geometry can reveal stable relations that reflect emerging functional organization. This motivates structure-aware regularization: rather than imposing cross-block coupling from the beginning, we activate it only after sufficiently organized block relations become observable, allowing the model to benefit from internal structure when such structure is likely to be informative.

To characterize this organization, we analyze block-wise similarity directly in parameter space. For each DiT block, we collect trainable square weight matrices, i.e., parameter tensors whose two dimensions are equal, and concatenate them into a block-level parameter representation. We then measure pairwise similarity between blocks using linear centered kernel alignment (CKA),
\begin{equation}
\mathrm{CKA}(w^i,w^j)
=
\frac{\mathrm{HSIC}(w^i,w^j)}
{\sqrt{\mathrm{HSIC}(w^i,w^i)\,\mathrm{HSIC}(w^j,w^j)}},
\end{equation}
where $\mathrm{HSIC}(\cdot,\cdot)$ denotes the Hilbert--Schmidt independence criterion under the linear kernel. Here, CKA is used as a parameter-space similarity measure for characterizing grouped organization across DiT blocks.

Empirically, we observe that block-wise relation patterns at early checkpoints are relatively diffuse, whereas later checkpoints exhibit clearer grouped organization under the same analysis protocol. Based on this observation, we construct a reference block-relation pattern, denoted by $S^{\mathrm{ref}}$, from the late-stage similarity structure used in our analysis. During training, we periodically estimate the current block-similarity matrix $S^{(k)}$ and use the reference pattern as a soft relational bias rather than a rigid target.

\begin{figure}[t]
    \centering
    \includegraphics[width=0.24\textwidth]{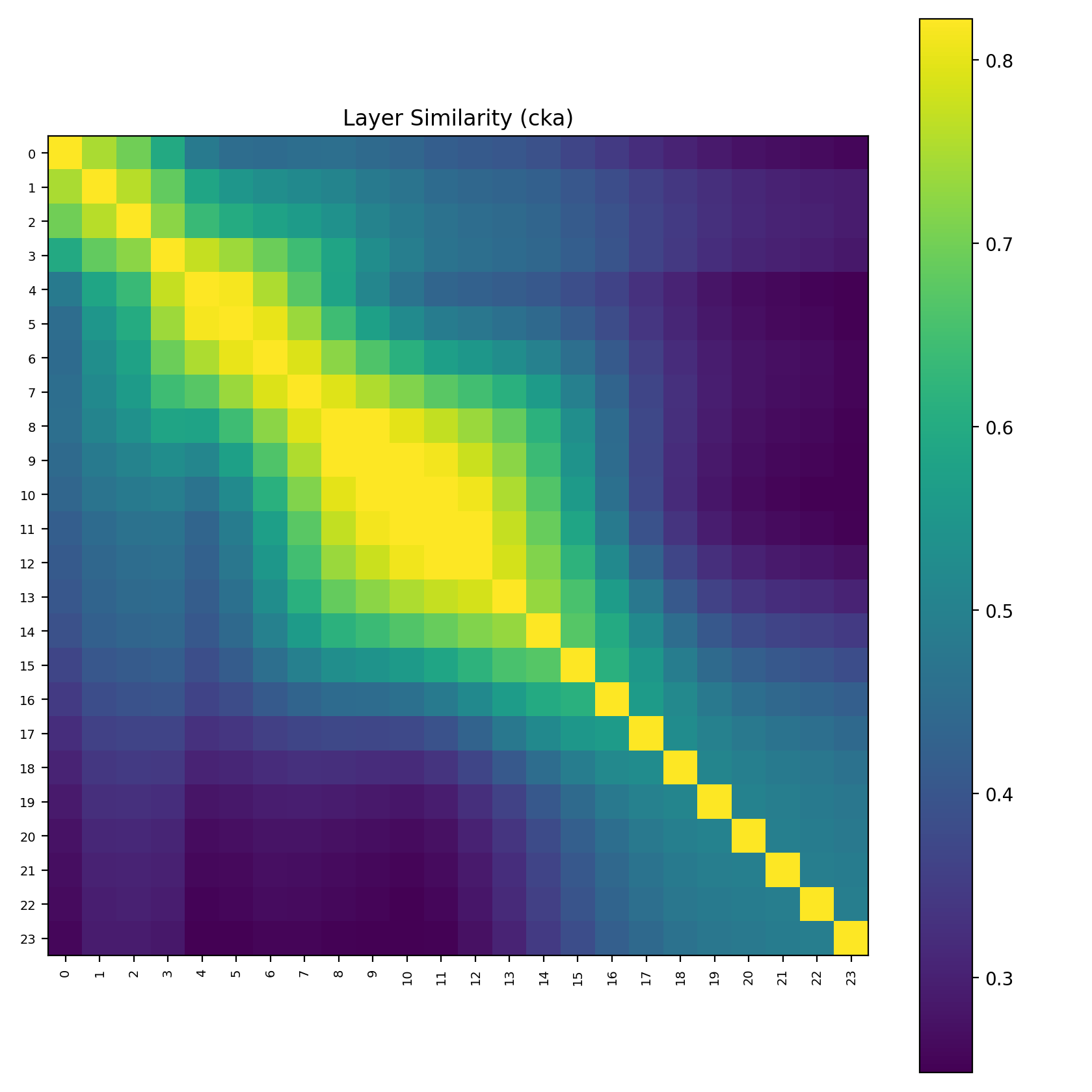}
    \hfill
    \includegraphics[width=0.24\textwidth]{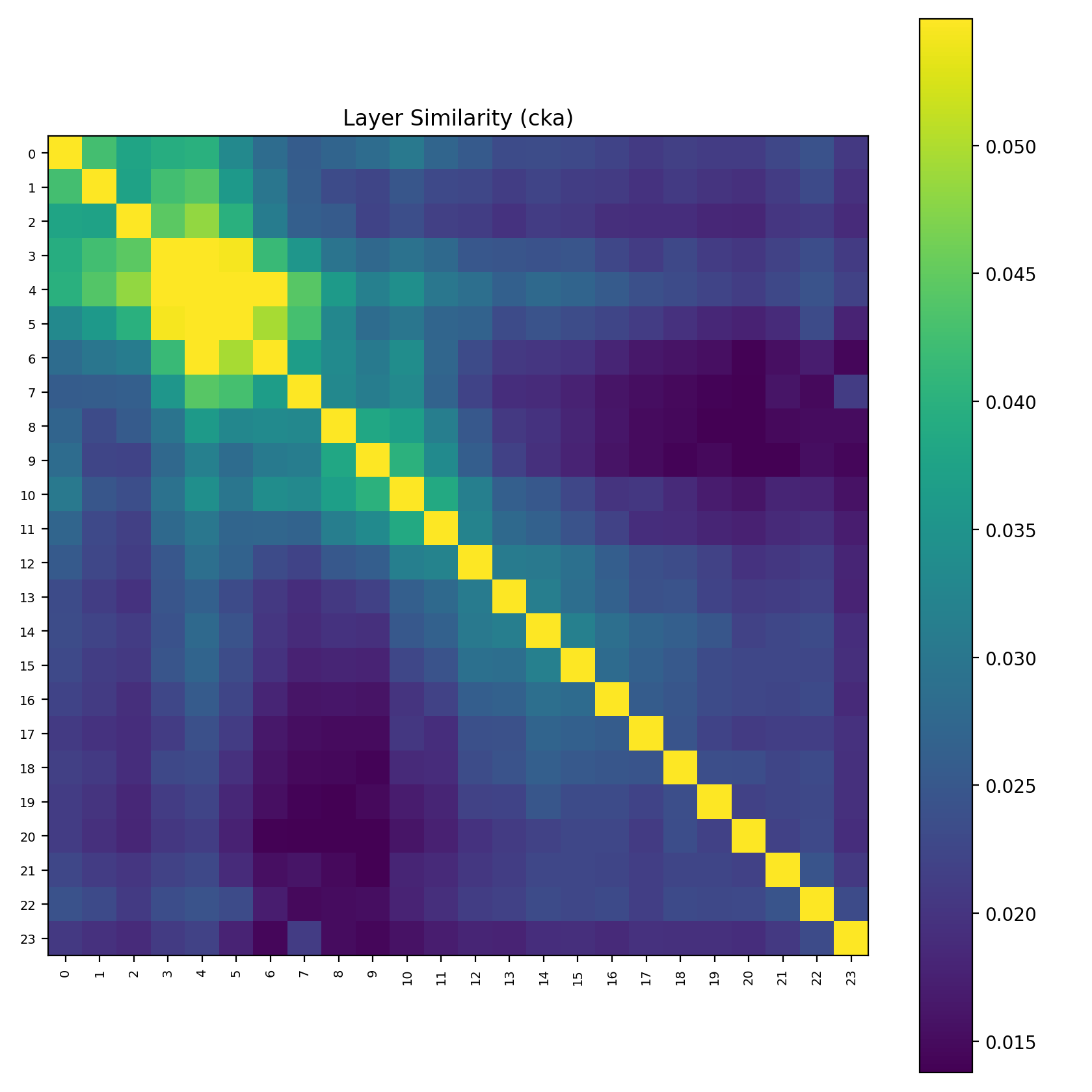}
    \hfill
    \includegraphics[width=0.24\textwidth]{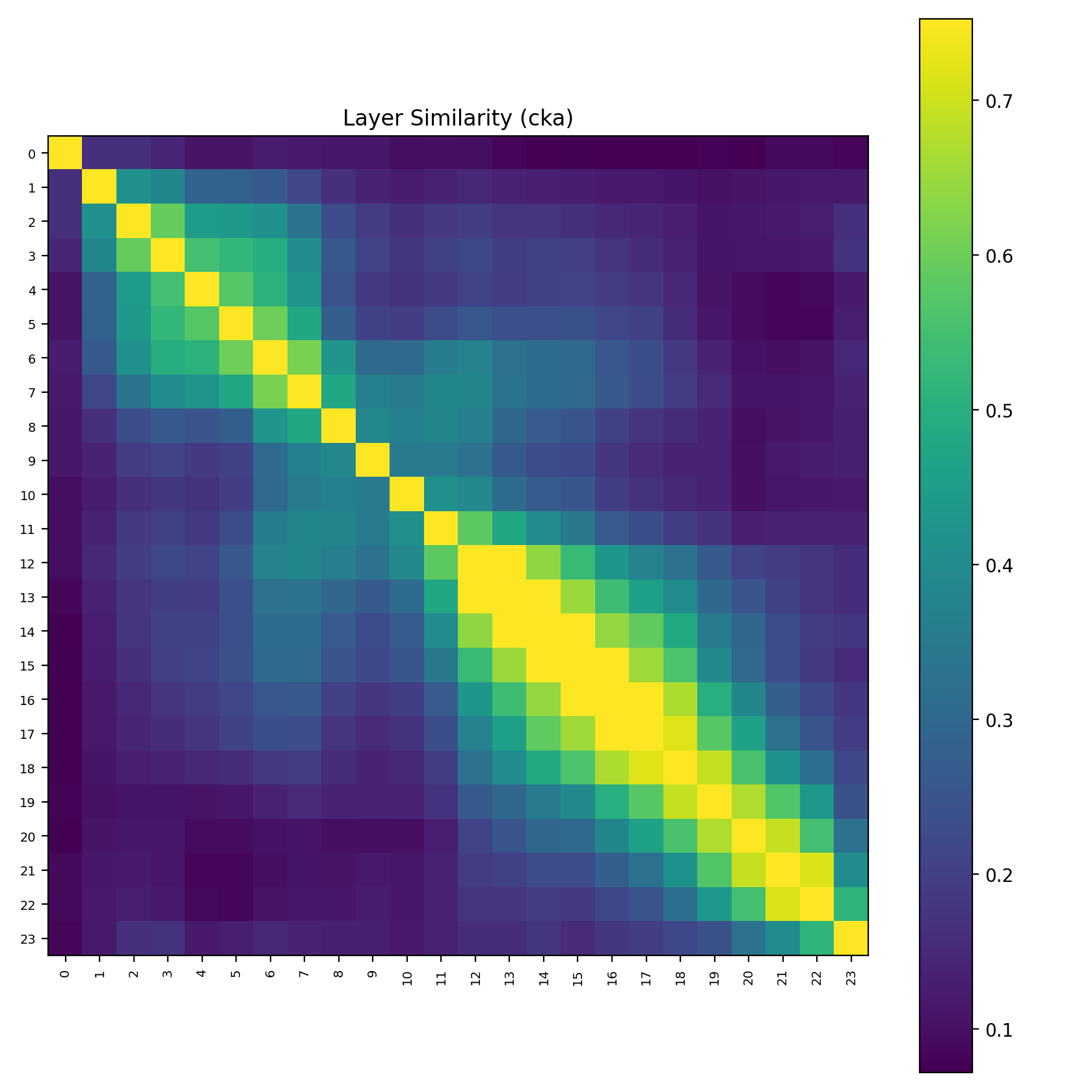}
    \hfill
    \includegraphics[width=0.24\textwidth]{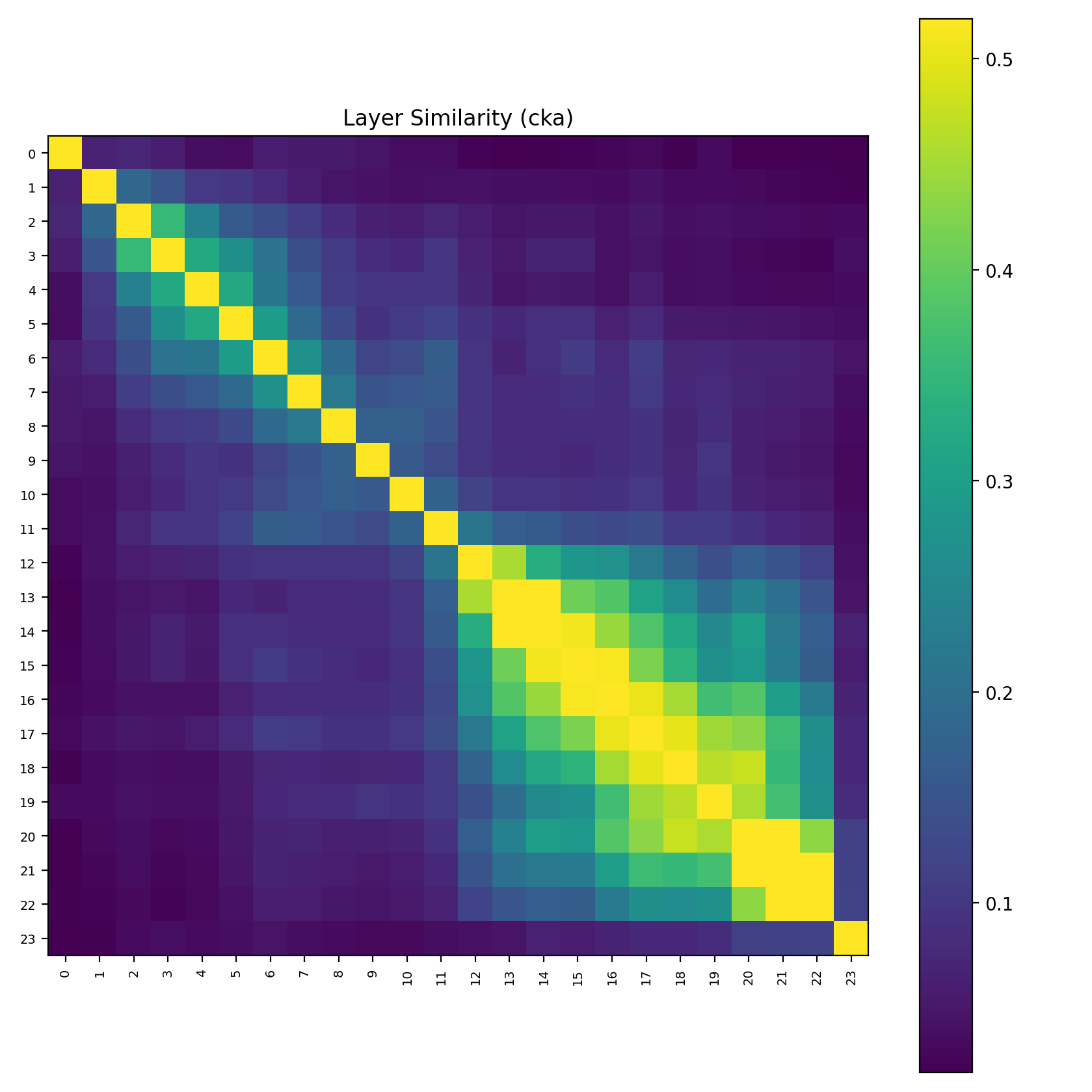}
    \caption{Evolution of block-wise parameter similarity during training. From left to right: 1K, 5K, 10K, and 100K steps. Similarity is computed from square weight matrices within each DiT block, providing an input-independent view of block coupling in parameter space. Early checkpoints exhibit relatively diffuse patterns, while clearer grouped organization gradually emerges as training proceeds.}
    \label{fig:cka_evolution}
\end{figure}

Figure~\ref{fig:cka_evolution} illustrates this trend. Block-wise similarity is relatively diffuse at early checkpoints, but develops clearer grouped organization as training proceeds. This suggests that structure-aware regularization should not be imposed uniformly from the beginning. Instead, cross-block coupling is introduced after internal relations become sufficiently organized, so that the regularizer acts as a refinement-stage inductive bias rather than an early constraint on representation formation.

From $S^{\mathrm{ref}}$, we construct a normalized affinity operator $W$ that encodes the reference coupling pattern among blocks. We then define a graph-smoothness regularizer over block parameters as
\begin{equation}
\mathcal{L}_{\mathrm{sp}}^{(k)}
=
\mathrm{Tr}\!\bigl((D-W)\,\Omega_k^\top \Omega_k \bigr),
\end{equation}
where $\Omega_k=[w_k^1;\dots;w_k^{L_b}]$ stacks the block-level parameter representations at iteration $k$, and $D$ is the degree matrix of $W$. This regularizer encourages blocks with stronger affinity under the reference relation pattern to evolve more coherently, while leaving weakly related blocks relatively unconstrained.

The strength of this regularizer is coupled to the same progress signal used throughout our framework. As introduced in Section~\ref{sec:overview_preliminaries}, $g_k$ measures the training-time rate of decrease of the SSL-space discrepancy and serves as a compact indicator of semantic progress. When $g_k$ is large, SSL-space progress remains rapid and strong structural regularization can be premature, since internal organization is still forming. As $g_k$ decreases, semantic progress becomes less pronounced, and organized block relations can provide a more useful internal bias for subsequent refinement.

Accordingly, we modulate the regularizer through a progress-dependent coefficient
\begin{equation}
\beta_k = \varphi(g_k),
\end{equation}
where $\varphi(\cdot)$ is a monotone decreasing map, so that the contribution of structure-aware regularization grows as semantic progress slows down. The overall training objective becomes
\begin{equation}
\mathcal{L}^{(k)}
=
\mathcal{L}_{\mathrm{diff}}^{(k)}
+
\lambda\,\beta_k\,\mathbf{1}(k \leq \rho_{\mathrm{sp}}K_{\mathrm{tot}})\,\mathcal{L}_{\mathrm{sp}}^{(k)},
\end{equation}
where $\lambda$ is the regularization weight, and $\rho_{\mathrm{sp}}K_{\mathrm{tot}}$ specifies the point after which the regularizer is removed to avoid interfering with final-stage detail refinement.

Under this design, the transition from external semantic guidance to internal structural regularization is guided by the same progress signal rather than by hand-designed phase boundaries. Early in training, optimization benefits more from external semantic support and the structural regularizer remains weak. As semantic progress slows and block relations become more organized, structure-aware regularization becomes increasingly active, providing an internal relational bias for refinement. In practice, we estimate $S^{(k)}$ only once every $M$ steps, which keeps the additional overhead low.
% =========================
% Experiments
% =========================
\section{Experiments}
\label{sec:experiments}

\subsection{Experiment Setup}
\label{sec:exp_setup}

We evaluate the proposed stage-aware mechanisms on two representative audio diffusion settings: text-conditioned audio generation and audio-conditioned audio super-resolution. For text-to-audio generation, we train on AudioSet~\cite{gemmeke2017audio} and FreeSound~\cite{fonseca2017freesound}, and evaluate on AudioCaps~\cite{kim2019audiocaps}. For audio super-resolution, we train on VCTK~\cite{yamagishi2019cstr} and evaluate on the VCTK-test split. We consider three reconstruction settings, namely 8-to-48~kHz, 16-to-48~kHz, and 24-to-48~kHz, where the model is conditioned on low-sample-rate audio and trained to recover the corresponding 48~kHz target.

All experiments are built on latent diffusion audio systems implemented in stable-audio-tools~\cite{evans2025stable}, with a Diffusion Transformer (DiT) as the denoiser. Within each task, we keep the backbone architecture, diffusion parameterization, and optimization budget fixed, and modify only the training mechanism under study. For SSL-guided and SSL-alignment components, we use a frozen pretrained USAD encoder as the audio self-supervised representation model throughout training~\cite{chang2025usad}. This setup isolates the effect of the proposed stage-aware mechanisms from confounding changes in model capacity or training scale.
\subsection{Baselines and Comparisons}
\label{sec:baselines}

For text-to-audio generation, we compare against two categories of baselines. The first is the default stable-audio-tools training pipeline, which serves as our uniform baseline. Starting from this baseline, we instantiate three stage-aware variants by introducing decayed SSL guidance, structure-aware regularization, and self-adaptive timestep sampling, respectively. We further compare against strong text-to-audio systems, including Make-An-Audio~\cite{huang2023make}, AudioLDM~2~\cite{liu2024audioldm}, and Tango~2~\cite{majumder2024tango}.

For audio super-resolution, we compare against two representative task-specific baselines, AudioSR~\cite{liu2024audiosr} and NVSR~\cite{liu2022neural}. Within our own framework, the uniform baseline is obtained by replacing the text condition in \texttt{stable-audio-tools} with low-sample-rate audio conditioning and training the model to reconstruct the corresponding 48~kHz waveform. We then evaluate the same three stage-aware variants under this setting, enabling a controlled comparison between static and stage-aware training strategies.

Taken together, these comparisons serve two purposes: first, to assess the contribution of each proposed mechanism within a common training framework; and second, to benchmark the resulting systems against strong task-specific baselines in both text-to-audio generation and audio super-resolution.
\subsection{Evaluation Metrics}
\label{sec:metrics}

We evaluate each method in terms of task-specific generation or reconstruction quality. For text-to-audio generation, we report Fr\'echet Audio Distance (FAD)~\cite{kilgour2018fr}, Kullback--Leibler divergence (KL), Inception Score (IS)~\cite{salimans2016improved}, and CLAP score~\cite{wu2023large}, capturing distributional fidelity, sample quality, and text-audio semantic alignment. For audio super-resolution, we report Log-Spectral Distance (LSD)~\cite{gray2003distance}, low-frequency LSD (LSD-LF), high-frequency LSD (LSD-HF), and SISNR~\cite{le2019sdr}, which together measure spectral reconstruction quality across different frequency bands as well as waveform-level fidelity.

Beyond task metrics, we use block-wise parameter similarity evolution to validate the stage-aware design of structure-aware regularization. This analysis is not intended as an additional benchmark metric, but as a mechanism-oriented diagnostic for examining when meaningful grouped organization begins to emerge during training.
\subsection{Main Results}
\label{sec:main_results}

\begin{table}[t]
    \centering
    \caption{Main results on text-to-audio generation.}
    \label{tab:main_results_tta}
    \begin{tabular}{lcccc}
        \toprule
        Method & FAD $\downarrow$ & KL $\downarrow$ & IS $\uparrow$ & CLAP $\uparrow$ \\
        \midrule
        Make-an-Audio 2 & 2.05 & 1.27 & -- & -- \\
        AudioLDM 2 & 1.86 & 1.64 & -- & -- \\
        Tango 2 & 2.69 & 1.12 & 9.09 & 0.57 \\
        Uniform baseline & 2.36 & 1.08 & 9.61 & 0.59 \\
        + Decayed SSL guidance & 2.08 & 1.04 & 10.67 & 0.59 \\
        + Structure-aware regularization & 2.12 & 1.06 & 10.16 & 0.62 \\
        + Self-adaptive timestep sampling & 1.91 & 1.04 & 10.92 & 0.62 \\
        \bottomrule
    \end{tabular}
\end{table}

\begin{table}
    \centering
    \caption{Main results on audio super-resolution over different input sampling rates. The target sampling rate is fixed at 48~kHz for all settings.}
    \label{tab:main_results_sr}
    \resizebox{\linewidth}{!}{
    \begin{tabular}{llcccc}
        \toprule
        Input SR & Method & LSD $\downarrow$ & LSD-LF $\downarrow$ & LSD-HF $\downarrow$ & SISNR $\uparrow$ \\
        \midrule
        \multirow{6}{*}{24 kHz}
        & AudioSR & 0.876 & 0.482 & 1.132 & 23.76 \\
        & NVSR & 0.845 & 0.451 & 1.104 & 22.14 \\
        & Uniform baseline & 0.831 & 0.445 & 1.098 & 22.51 \\
        & Decayed SSL guidance & 0.760 & 0.429 & 1.060 & 22.68 \\
        & Structure-aware regularization & 0.772 & 0.427 & 1.048 & 22.27 \\
        & Self-adaptive timestep sampling & 0.769 & 0.423 & 1.043 & 22.53 \\
        \midrule
        \multirow{6}{*}{16 kHz}
        & AudioSR & 1.108 & 0.473 & 1.307 & 18.71 \\
        & NVSR & 0.863 & 0.232 & 1.042 & 18.53 \\
        & Uniform baseline & 0.878 & 0.231 & 1.079 & 19.17 \\
        & Decayed SSL guidance & 0.843 & 0.219 & 1.053 & 19.54 \\
        & Structure-aware regularization & 0.813 & 0.217 & 1.047 & 19.18 \\
        & Self-adaptive timestep sampling & 0.838 & 0.206 & 1.049 & 19.09 \\
        \midrule
        \multirow{6}{*}{8 kHz}
        & AudioSR & 1.271 & 0.383 & 1.379 & 12.97 \\
        & NVSR & 1.018 & 0.370 & 1.102 & 12.97 \\
        & Uniform baseline & 1.134 & 0.376 & 1.487 & 12.73 \\
        & Decayed SSL guidance & 1.029 & 0.349 & 1.221 & 12.40 \\
        & Structure-aware regularization & 1.014 & 0.342 & 1.196 & 12.34 \\
        & Self-adaptive timestep sampling & 1.021 & 0.341 & 1.170 & 12.92 \\
        \bottomrule
    \end{tabular}
    }
\end{table}

Tables~\ref{tab:main_results_tta} and~\ref{tab:main_results_sr} summarize the main results on text-to-audio generation and audio super-resolution. Across both settings, the proposed stage-aware mechanisms consistently outperform the uniform baseline in final quality, training efficiency, or both, supporting the view that audio diffusion training benefits from stage-dependent optimization rather than a fixed recipe. One exception is SISNR in audio super-resolution, where the gains are less pronounced than those observed on spectral metrics such as LSD, LSD-LF, and LSD-HF. We speculate that this gap may arise from a mismatch between the semantic structure emphasized in latent-space learning and the waveform-level fidelity emphasized by SISNR, such that improvements in the learned latent representation do not necessarily translate into equally strong gains under waveform-space reconstruction measures.

Figure~\ref{fig:cka_evolution} provides further evidence for the gated design of structure-aware regularization. Block-wise parameter similarity is relatively diffuse at early checkpoints and gradually develops clearer grouped organization as training proceeds. The same qualitative trend appears in all 28-blocks, 24-block and 20-block DiTs, and similar final relation patterns are observed across multiple data distributions in separate small-scale analyses. Together, these results suggest that grouped internal organization is neither depth-specific nor dataset-specific, but instead reflects a more general structural tendency of diffusion training. This observation supports activating structure-aware regularization only after meaningful internal organization has formed.

\section{Related Works}
\label{sec:related_works}

\paragraph{Diffusion models.}
Diffusion probabilistic models have become a dominant paradigm for generative modeling by learning to reverse a gradual noising process~\cite{ho2020denoising,song2020score,karras2022elucidating}. They have achieved strong performance in image generation, video generation, and audio generation, driven by advances in scalable architectures, conditioning mechanisms, and improved training objectives~\cite{ho2022imagen,singer2022make}. In visual generation, diffusion backbones have evolved from convolutional latent U-Net architectures to transformer-based architectures and masked diffusion transformers, substantially improving model scalability and contextual modeling~\cite{rombach2022high,peebles2023scalable,gao2023masked}. In audio generation, diffusion models have also shown strong performance across heterogeneous conditioning regimes, including text-conditioned audio generation, multimodal audio synthesis, and quality-aware music generation~\cite{kreuk2022audiogen,li2024quality}. These systems provide strong foundations for conditional audio diffusion, while also motivating closer study of how semantic guidance, timestep allocation, and structural constraints should be organized throughout optimization.

\paragraph{Efficient diffusion training.}
Recent work has improved diffusion training efficiency from complementary perspectives. Representation-based methods regularize denoising-network features using external semantic representations, structural constraints, or internal representation signals, improving optimization efficiency and generation quality~\cite{leng2025repa,chen2025structural,wu2025representation,jiang2025no,wang2025diffuse}. Orthogonal to representation learning, timestep-based methods show that the training contribution of different noise levels is highly non-uniform, and improve efficiency by modifying objectives, importance weighting, or timestep allocation~\cite{nichol2021improved,wang2025closer}. These studies suggest that semantic representation, internal structure, and timestep allocation are important factors in diffusion training. Building on these observations, our work studies audio diffusion training under a unified stage-aware view, where semantic support, structural coupling, and timestep emphasis are jointly adapted by a shared progress signal derived from frozen audio SSL representations.
% =========================
% Conclusion
% =========================
\section{Conclusion}
\label{sec:conclusion}

We presented a stage-aware perspective on efficient audio diffusion training across heterogeneous conditioning regimes. Instead of treating training as a static optimization process, we argued that external guidance, internal organization, and noise-level allocation should play different roles at different stages of learning. Motivated by this view, we studied three complementary mechanisms: decayed SSL guidance, structure-aware regularization based on emergent grouped organization in parameter space, and self-adaptive timestep sampling driven by a global SSL-alignment signal.

Experiments on text-conditioned audio generation and audio-conditioned super-resolution show that these stage-aware mechanisms improve generation quality and training efficiency over standard static training recipes. In particular, the results suggest that external semantic priors are most useful as an early semantic scaffold, grouped internal organization becomes informative only after it has emerged, and timestep allocation should adapt once semantic alignment begins to saturate. Beyond the quantitative gains, the observed structural evolution during training provides empirical support for the view that audio diffusion optimization is inherently stage-dependent.

More broadly, our findings suggest that efficient audio diffusion training is better understood as a progressive process in which semantic acquisition, structural organization, and detail refinement should not be optimized with a single fixed recipe. We hope this perspective will motivate future work on training-efficient generative modeling beyond the specific mechanisms studied here.

\bibliographystyle{plainnat}
\bibliography{references}
\end{document}